\documentclass[preprint,nofootinbib]{revtex4}

\usepackage{amsmath,amssymb}

\usepackage{epsfig}
\usepackage{color}

\newcommand{\n}{\nonumber}
\newcommand{\be}{\begin{equation}}
\newcommand{\ee}{\end{equation}}
\newcommand{\bea}{\begin{eqnarray}}
\newcommand{\eea}{\end{eqnarray}}

\begin{document}

\title{Effective Free Energy Landscapes and Black Hole Thermodynamic Phase Transitions}
\author{Choon-Lin Ho}%\footnote{email: hcl@mail.tku.edu.tw}}%\\
\affiliation{Department of Physics, Tamkang University,
Tamsui 25137, Taiwan}
% Phys. Scr. 101 (2026) 025006

%(WS)
%\affiliation

%\date{2024/2/8} 

%\maketitle  % for WS

\begin{abstract}

A recent interesting development in the dynamics of black hole phase transitions has been the so-called Gibbs free energy landscape approach.    In this formalism, it is assumed  that there exists a canonical ensemble of a series of black hole spacetimes with arbitrary horizon radius at a given ensemble temperature.   An off-shell Gibbs free energy  is defined for every spacetime state in the ensemble, with the horizon radius treated as the order parameter.   The minima (maxima) of this function correspond to the various stable (unstable)  black hole states.   This off-shell Gibbs free energy is then treated as a classical effective drift potential  of  an associated Fokker-Planck equation used to study the stochastic dynamics of black hole phase transition under thermal fluctuations.  Additive noise, which is independent of the black hole size,  is assumed in obtaining the Fokker-Planck equation.   In this work we extend the previous treatment by considering the effects of multiplicative noise, namely, noise that could scale with black hole size.  This leads to an effective free energy function that can be used to study the modification of the thermodynamic phase transition of a black hole system.  
It is realized that it is generally difficult to form black holes under a multiplicative noise, 
 unless the effective and the original free energy become extremal at the same horizon radius.  For this latter situation some
theoretical noise profiles which are  monotonically increasing/deceasing functions of the horizon radius are considered.  It is found that stronger noise disfavors the formation of black hole.

\end{abstract}

\maketitle
%%%%%%%%%%%%%%%%%
\section{Introduction}
In the early seventies of the last century,  Bekenstein and Hawking revealed the thermal nature of black holes by associating a black hole with an entropy \cite{ B} and a temperature \cite{H}. Since then black hole thermodynamics \cite{BBH} has been an exciting and intriguing area of research in black hole physics that establishes a deep link between gravity, thermodynamics, and quantum physics (for reviews, see eg., \cite{W,C}).

Among various aspects of black hole thermodynamics, the study of  structures and transitions of  thermodynamic phases of some black hole systems has received great interest in recent years (see eg. [6-16] for some works on various aspects of such study in different types of black hole systems).  Hawking-Page phase transition was  found for the  Schwarzschild-AdS (anti-de Sitter) black hole system, which is a first-order transition between a thermal AdS space and  the large AdS black hole at a certain critical temperature \cite{HP}. The first-order phase transition between small and large black holes was studied for the charged Reissner-Nordstr\"om (RN)-AdS black hole \cite{CEJM,Wu}.  
 Further study reveals that the cosmological constant should play the role of a thermodynamical state variable \cite{S}, and soon it was treated as a pressure in the equation of state of black hole thermodynamics \cite{D1,K2}.  The  van der Waals liquid-gas type phase transitions were later studied for the charged black holes \cite{M1} and the charged rotating black holes system \cite{D2}.
   These interesting results soon inspired various studies of phase structures and transitions of other black hole systems (for a review, see eg., \cite{M2}).  

A recent interesting development on the dynamics of black hole phase transitions has been the so-called free energy landscape approach to the Hawking-Page transition considered in \cite{LW1}.  In this formalism, it is assumed  that there exists a canonical ensemble of a series of black hole spacetimes with arbitrary horizon radius at a given ensemble temperature.  This ensemble consists also intermediate states which are not solutions of the Einstein equation as well as stable and unstable black hole solutions.    An off-shell Gibbs free energy, obtained by replacing the Hawking temperature of the on-shell free energy by the ensemble temperature,  is defined for every spacetime state in the ensemble.  
The horizon radius is treated as the order parameter.
 The minima (maxima) of this function correspond to the various stable (unstable)  black hole states.   This generalized Gibbs free energy is then treated as an effective drift potential  of  an associated Fokker-Planck equation used to study the stochastic dynamics of black hole phase transition under thermal fluctuations.   This approach has been applied to phase dynamics of the the Schwarzschild-AdS systems in \cite{LW1}, and to the RN-AdS systems in \cite{LW2}.  
 It was later shown that the generalized free energy can be derived from the Einstein-Hilbert action of the Euclidean gravitational instanton with a conical singularity \cite{LW3}.
Subsequently, other forms  of free energy landscapes have been proposed, such as the Landau free energy \cite{X1} and the thermal potential \cite{X2}.   
The landscape approaches have since been extended to other black holes systems, eg.,
the Kerr-AdS black hole \cite{Y}, Hawking-Page transition with non-Markovian effects \cite{LW4}, the dyonic AdS black hole \cite{LW5}, black hole in massive gravity \cite{WLL}, etc.
Kramers escape rates of phase transitions have also been  considered recently \cite{X3,A1,X4,A2}.

So far the study of the stochastic dynamics of the black hole phase transitions is based mainly on uniform additive thermal noise which is independent of the horizon radius.  However,  in real situations it is not unreasonable to assume that the noise would depend on the size of black hole, considering the fact that thermal,  quantum, and spacetime fluctuations near  black holes of different sizes could be different.
In this work we would like to explore the effect  on the thermodynamic phase transition of black hole states if the noise could scale with space.  Such noise is called multiplicative noise. 

\smallskip
%%%%%%%%%%%%%%%%%%%%%%%
\section{Fokker-Planck\,equation}
Consider a four-dimensional Schwarzschild-AdS black hole with mass $M$ and AdS curvature radius $L=\sqrt{-3/\Lambda}$, where $\Lambda$ is the cosmological constant. The metric (in $G=1$ units) is \cite{LW1}
\bea
ds^2 &=& -f(r) dt^2 + f(r)^{-1}\,dr^2 +r^2 \left(d\theta^2+\sin^2\theta d\phi^2\right),\n\\
f(r) &=&1-\frac{2M}{r}+\frac{r^2}{L^2}.
\label{ds}
\eea
Setting $f(r)$ to zero gives the black hole horizon radius $r_+$, which in this case there is only one real solution.
In terms of $r_+$, the mass $M$, the Hawking temperature $T_H$,  and the Bekenstein-Hawking entropy $S$ are
\be
M=\frac{r_+}{2}\left(1+\frac{r_+^2}{L^2}\right),~T_H=\frac{1}{4\pi r_+}\left(1+\frac{3r_+^2}{L^2}\right),~
S=\pi r_+^2.
\label{data}
\ee
The expression of $T_H$ implies that a Schwarzschild-AdS black hole has a minimal temperature (c.f. Fig.\,1)
\be
\mathcal{T}_m=\frac{\sqrt{3}}{2\pi L}.
\label{Tm}
\ee

%%%%%%%%%

In  \cite{LW1} it was proposed to consider Hawking-Page transition in the so-called free energy landscape formalism. In such formalism,  one considers a canonical ensemble  of a series of black hole spacetimes with arbitrary horizon radius  at temperature $\cal T$. Phase transition is then analyzed based on the  Gibbs free energy defined for every spacetime state.  The horizon radius $r_+$  is treated as an order parameter. 
The Gibbs free energy for the Schwarzschild-AdS black hole is given by
$\mathcal{G}=M-{\cal T}_H S$.    As the ensemble consists also intermediate states which are not solutions of the Einstein equation, a so-called off-shell  Gibbs function for the ensemble is constructed by replacing the Hawking temperature $\cal{T}_H$  by the  the ensemble temperature $\cal T$, i.e., $\mathcal{G}=M-{\cal T} S$, which in this case is
\be
\mathcal{G}(r_+, \mathcal{T}) = \frac{r_+}{2}\left(1+\frac{r_+^2}{L^2}\right)-\pi {\mathcal T} r_+^2,
\ee
or, in terms of the two dimensionless variables $r\equiv r_+/L$ and $T\equiv {\mathcal T}L$,
\bea
\mathcal{G}(r_+, \mathcal{T}) &=& L\, G_0(r,T),\label{G0}\\
G_0(r,T) &=& \frac{r}{2}\left(1+ r^2\right)-\pi T r^2.\n
\eea
The minima (maxima) of this function correspond to the various stable (unstable)  black hole states, as $G_0^\prime=0$ gives the relation between $T_H$ and $r_+$ in (\ref{data}).

The Hawking-Page transition is easily understood in this landscape picture using $G_0(r,T)$, as shown in Fig.\,2.  There are two critical temperatures: the minimal temperature $T_m$ in (\ref{Tm}), where $G_0$ exhibits an inflection point at $r=1/\sqrt{3}$, and the Hawking-Page temperature $T_{HP}=1/\pi $, where $G_0$ has two degenerate global minima at $r=0, 1$. For $T<T_m$, there is just one global minimum of $G_0$ at $r=0$, representing the system is in a pure radiation phase, or the thermal AdS space. When $T_m<T<T_{HP}$, a local maximum  and a
local minimum  appear for  $r>0$, corresponding to an unstable small black hole phase and a metastable large black hole phase, respectively.  For $T>T_{HP}$, the large black hole phase is the stable state.   

The stochastic kinetics of the states in a thermodynamic ensemble under thermal fluctuation can be developed  in terms of a Langevin equation with $\mathcal{G}$ as the external force potential, and a stochastic noise $\xi(t)$, which in \cite{LW1} is assumed implicitly to be a space-independent Gaussian noise,
\be
\dot{r}_+(\mathcal{T},t)= -\frac{\mathcal{G}^\prime (r_+,\mathcal{T})}{\zeta} + \xi(t).
\label{LE}
\ee
Here $\zeta$ is the dissipation coefficient, the ``dot" and the ``prime" represent derivatives with respect to time $t$ and space $r_+$, respectively, and 
\be 
\langle \xi(t)\rangle =0,~~ \langle \xi(t)\xi(t^\prime)\rangle= 2 D \delta(t-t^\prime),~~~D\equiv \frac{k \mathcal{T}}{\zeta},
\ee
where $k$ is the Boltzmann constant.
The Fokker-Planck equation corresponding to (\ref{LE}) is \cite{R}
\be
\frac{\partial }{\partial t}P(r_+,\mathcal{T},t)= \frac{\partial}{\partial r_+}\left(\frac{\mathcal{G}^\prime (r_+,\mathcal{T})}{\zeta}\,P(r_+,\mathcal{T},t)\right)
+ D  \frac{\partial^2}{\partial r_+^2}\,P(r_+,\mathcal{T},t).
\label{FP}
\ee
Here $P(r_+,\mathcal{T},t)$ is the probability density function of the states in the thermal ensemble.

Using (\ref{FP}), one can then study various aspects of the stochastic kinetics of the thermal ensemble of the black hole states, such as  the mean first passage time for the black hole state switching and Hawking-Page transition \cite{LW1}. 
 Particularly, the stationary state $P_0(r_+,\mathcal{T})$ is given by $\partial P/\partial t=0$ with the boundary conditions $P_0(r_+,\mathcal{T})\to 0$ as $r_+\to 0, \infty$. 
 Integrating
\be
\frac{\partial }{\partial t}P(r_+,\mathcal{T},t)=0 \Longrightarrow  
D \frac{\partial}{\partial r_+} P_0(r_+, \mathcal{T})=-\frac{\mathcal{G}^\prime (r_+, \mathcal{T})}{\zeta}\,P_0(r_+, \mathcal{T})\ee
gives
\bea
P_0(r_+,\mathcal{T}) &\sim& \exp\left(-\frac{\mathcal{G}(r_+,\mathcal{T})/\zeta}{D}\right) \label{P0}\\
&\sim& \exp\left(-\frac{\mathcal{G}(r_+,\mathcal{T})}{k\mathcal{T}}\right).\n
\eea
Thus the global minimum of $\mathcal{G}(r_+,\mathcal{T})$ gives the largest probility density, representing the most likely black hole spacetime in the canonical ensemble.

\smallskip
%%%%%%%%
\section{Effective free energy landscape}
In (\ref{LE}) the noise is additive, as it is space-independent and added directly to the equation governing the change in $r$.  
But in real situations it is not unreasonable to assume that the noise would depend on the radius  of the horizon, considering the fact that thermal,  quantum, and spacetime fluctuations near  black holes of different sizes could be different.
As such, we would like to explore the effect  on the Hawking-Page transition if the noise could scale with space.  Such noise is called multiplicative noise.

The general Langevin equation with such a noise force is \cite{R}
\be
\dot{r_+}(\mathcal{T},t)= -\frac{\mathcal{G}^\prime (r_+,\mathcal{T})}{\zeta} + g\left(\frac{r_+}{L}\right))\xi(t).
\label{GLE}
\ee
Here the noise profile $g(\cdot)\neq 0$ is a dimensionless scaling function of the Gaussian noise, which for simplicity  we assume  to be time and temperature independent. 
 It tells how the strength of the stochastic white noise $\xi(t)$ varies for  different $r_+$.  The additive noise in (\ref{LE}) corresponds to $g=1$,  which acts equally on the black hole states with different $r_+$. 
 
It is well known that there are two well-defined rules to integrate a stochastic differential equation, such as (\ref{LE}) and (\ref{GLE}), namely, the It\^o's  and  the Stratronovich's rule (please see for example \cite{R} for details).  In physical applications which rule to adopt depends on which one best describes the phenomenon one wishes to model.   Accordingly, the Fokker-Planck equation associated with the generalized Langevin equation will be slightly different.  For our present case,  the Fokker-Planck equation corresponding to 
(\ref{GLE}) is \cite{R}
\be
\frac{\partial }{\partial t}P= \frac{\partial}{\partial r_+}\left[\left(\frac{\mathcal{G}^\prime }{\zeta}-\lambda\,Dgg^\prime\right)P\right]
+ \frac{\partial^2}{\partial r_+^2}\left(Dg^2 P\right).
\label{FP1}
\ee
Here the parameter $\lambda=0, 1$ according to the It\^o's  and Stratronovich's rule, respectively.
Eq.\,(\ref{FP1}) reduces to (\ref{FP}) for $g=1$, i.e., both the It\^o and the Stratronovich's rule give the same Fokker-Planck equation for additive noise.

As before, the stationary solution $P_0(r_+,\mathcal{T})$ of (\ref{FP1}) is given by $\partial P/\partial t=0$ and the boundary conditions $P_0(r_+,\mathcal{T})\to 0$ as $r_+\to 0, \infty$. This is given by the solution of
\bea
\frac{\partial}{\partial r_+}\left(Dg^2 P_0\right)&=&-\left(\frac{\mathcal{G}^\prime }{\zeta}-\lambda\,Dgg^\prime\right)P_0\\
&=&-\left[ \left(\frac{\mathcal{G}^\prime }{\zeta}-\lambda\,Dgg^\prime\right)/(Dg^2)\right]\left(Dg^2 P_0\right),\n
\eea
which is easily integrated to give
\bea
P_0(r_+,\mathcal{T})&\sim& \exp\left(-\frac{G(r_+,\mathcal{T})}{k\mathcal{T}}\right),\\
G(r_+, \cal{T})&\equiv& \int^{r_+}_c\,\frac{\mathcal{G}^\prime}{g^2}dr_+ + k\mathcal{T}\left(2-\lambda\right) \ln g,\n
\eea
where $c$ is a constant.
Comparing with (\ref{P0}), this solution is the same as the stationary state of a Fokker-Planck equation (\ref{FP}) with $G(r_+,\mathcal{T})$ as the drift potential.  Thus the phase structure of  the black hole states in the thermodynamic ensemble with multiplicative noise can be equivalently 
studied as that with an additive noise, but with $G$ as the effective Gibbs free energy landscape. 
In terms of the dimensionless variables, $G(r_+,\mathcal{T})$ is
\bea
G(r,T) &=& \int^r_c\,\frac{1}{g^2}\left(\frac32 r^2-2\pi T r +\frac12 \right)dr \n\\
&&+ \frac{k T}{L^2} \left(2-\lambda\right) \ln g(r).
\label{Ge}
\eea

For a uniform additive noise $g(r)=1$, we have $G(r,T)=G_0(r,T)$. So the classical landscape remains unchanged. This is not the case for general noise profile.   Particularly, unlike the case with $G_0(r, T)$, the extrema of $G(r,T)$ do not necessarily correspond to the black holes solutions of the Einstein equations for general noise, unless $G^\prime (r,T)=0$ and $G_0^\prime (r,T)=0$ share same common roots at a given temperature $T$ (at these common roots we have $g^\prime (r)=0$). Thus it is seen that it is generally difficult to form black holes under a multiplicative noise.

If  $G(r,T)$ has the same functional form as $G_0(r,T)$ in some part of the half-line in which $G_0(r,T)$ has extrema, then $G(r,T)$ admits black hole solutions.   For instance, one might consider the situation where thermal fluctuations differ significantly from additive white noise only for black hole states with small horizon radii.  This is not unreasonable in view of the strong gravitational and quantum effects near small black holes. 

Here we would like to consider one such theoretical noise profile to
see how multiplicative noise could affect the nature of the Hawking-Page transition of a Schwarzschild-AdS black hole systems.    For definiteness, we adopt the Stratronovich's rule ($\lambda=1$) and set $k=L=1$.  The It\^o's rule ($\lambda=0$)  gives qualitatively similar results. 

We take $g(r)=1+a - sign(a) \sqrt{a^2-(r-|a|)^2}$ for $r<|a|$, and $g(r)=1$ for $r>|a|$ ($a>-1$).   The noise differs from the uniform additive noise $g(r)=1$ only in $r<|a|$, and it is stronger (weaker) than the additive white noise  for $a>0$ ($-1<a<0$).  For $a>0$, $g(r)$ decreases from $g(0)=1+a$ to $g(a)=1$ along a circular arc of radius $a$. For $a<0$, $g(r)$ increases from $g(0)=1-|a|$ to $g(a)=1$ along a circular arc.  With this profile, black hole spacetimes with smaller horizons experience greater changes of noise level from the the additive noise in (\ref{G0}).  

Fig.\,3 shows $G(r, T_{HP})$ for different values of $a$ at $T_{HP}=1/\pi$, the Hawking-Page critical temperature in the case of additive noise (i.e., $a=0$).  The constant $c$ is chosen such that $G(0,T)=0$.   It is seen that for $-1<a <0$ (weaker noise level near $r=0$), the system is still in the thermal AdS state, and for $a>0$ (stronger noise level near $r=0$), the system is already in the large black hole state.  One can understand this effect as weaker  noise fluctuation near  $r=0$ favors the formation of the thermal AdS states and disfavors the formation of the large black hole.   Thus one would expect a higher (lower) critical temperature for phase transition  for $-1<a<0$ ($a>0$).   It is worthy to note that for $a>0$, $G(r,T)$  could develop a new local minimum in $r<a$. 
 However, this minimum does not correspond to a solution of the Einstein equation, as the values of $r$ and $T$ at this minimum do not satisfy the Hawking  temperature relation in (\ref{data}). In the original free  energy landscape it represents a transient state \cite{LW1}.  Here we shall call such states the metastable transient states.
 
 For $a<0$ the phase structures of the thermodynamic black hole spacetimes are qualitatively similar to that of the case with $a=0$.
However,  for cases with $a>0$,
it is anticipated that the possible presence of the new minimum  will add new features to the phase structure of the thermodynamic black hole spacetimes.  As a representative example  we show in Fig.\,4 the effective free energy $G(r, T)$ as a function $r$ for the case with $a=0.2$ at different temperatures.   It shows that already at $T_m$, a local minimum corresponding to the large black hole state begins to appear, while for $a=0$ case this state is degenerate with the unstable medium black hole state at the inflection point.  As $T$ increases, the large black hole state gradually becomes the stable state, and the local minimum, i.e., the meta stable transient state, remains in the $g(r)>1$ (i.e., $a<0.2$) region before it disappears at sufficiently high $T$.

The phase structures of the black hole states may be understood more directly in terms of the plots of $G(r,T)$ vs $T$.  Fig.\,5  presents such plots  for both $a=0$ and $0.2$.  The intersection of the curves for  the large black hole states and the thermal AdS states/transient states give the phase transition temperatures, which is seen to be lower for $a=0.2$ than that for $a=0$.  This indicates that large black hole states form earlier for $a>0$ as temperature increases, as mentioned before.  For $a=0$, the curves for the unstable black hole states approaches $G=0$ (for the thermal AdS states)  asympotically.  But for $a=0.2$, the curves for the unstable black hole states and the metastable transient states meet at  $T\approx 0.45$.  These two kinds of states become degenerate at this temperature.  For $a<0$, the phase diagram is similar to that of $a=0$.

\smallskip
%%%%%%%%%%%%%%%
\section{RN-AdS black hole}
We now extend the previous consideration to the corresponding situation for the RN-AdS system as discussed in \cite{LW2}. 

Following \cite{LW2}, the basic data of RN-AdS system are summarized below.
The metric in this case is given by (\ref{ds}) but with $f(r)$ changed to
\be
f(r) =1-\frac{2M}{r}+\frac{r^2}{L^2}+ \frac{Q^2}{r^2},
\ee
where $Q$ is the charge of the black hole.
The Schwarzschild-AdS black hole can be viewed as the $Q\to 0$ limit of the RN-AdS system.

 The largest root of the equation $f(r)=0$ gives the radius $r_+$ of the event horizon.
The mass $M$ and the Hawking temperature $T_H$ are
\bea
M &=& \frac{r_+}{2}\left(1+\frac{r_+^2}{L^2}+ \frac{Q^2}{r_+^2}\right),\\
\mathcal{T}_H &=& \frac{1}{4\pi r_+}\left(1+\frac{3 r_+^2}{L^2} - \frac{Q^2}{r_+^2}\right).
\eea

In \cite{LW2} thermal phase structure of the RN-AdS black hole system at different ensemble temperature $\mathcal{T}$  is studied by treating the cosmological constant $L$  (in terms of the related thermal pressure $P=3/8\pi L^2$ \cite{D1}) as a control parameter, keeping the charge $Q$ fixed.   However,  in order to be in line with the discussion in the previous sections, we will  keep $L$ fixed 
(and set $L=1$ for numerical analysis) and treat $Q$ as the control parameter.  Thus we will rewrite some relevant formulas in \cite{LW2} for our purpose.

In terms of the dimensionless variables $r=r_+/L, T=\mathcal{T}L$ and $q=Q/L$,
The Hawking temperature can be re-expressed as
\be
T_H=\frac{1}{4\pi r}\left(1+ 3 r^2 - \frac{q^2}{r^2}\right).
\label{Th}
\ee
$T_H$ is a monotonic function of $r$ when $q>q_c=1/6$, and has a local minimum $T_m$ and a local maximum $T_M$ otherwise.  Fig.\,6 depicts these situations for $q=1/3$ and $1/10$. 

It is clear that there is no phase transition if $q>q_c$.  For $q<q_c$,  $T_m$ and $T_M$ are
given by
\be
T_{m/M}=\frac{1}{\pi}\sqrt{\frac32}\frac{\left(1- 12 q^2 \pm \sqrt{1-36 q^2}\right)}{\left(1\pm \sqrt{1-36 q^2}\right)^{3/2}}.
\ee
Note that in the limit $q\to 0$ (the Schwarzschild-AdS limit), $T_M$ does not exist and $T_m$ reduces to (\ref{Tm}) (in dimensionless form).
For $T_m< T_H< T_M$ there exists three branches of black hole solutions: small, large and (unstable) intermediate black holes. There can be a first-order van der Waals phase transition between the small and large black holes. 

Again one considers a canonical ensemble at a temperature $T$ consisting of a series of black hole spacetimes with arbitrary horizon radius.  Spacetimes other than the three branches of the black holes are off-shell, i.e., they are not solutions of the Einstein equation.  The off-shell Gibbs free energy function is
\be
G_0(r, T)=\frac{r}{2}\left(1+ r^2 +\frac{q^2}{r^2}\right)-\pi T r^2.
\label{G2}
\ee
In Fig.\,7 we give the graphs of $G_0(r,T)$ vs $r$ with $q=1/10$ for different $T$. Here $T_m=0.27120$ and $T_M=0.34869$. For $T<T_m$,  $G_0$ has one global minimum corresponding to the small black hole. An inflection point occurs at $T_m$.  For $T$ in between  $T_m$ and $T_M$ there are two local minima and a local maximum corresponding to the three black holes solutions mentioned before.  When $T>T_M$, there is only one global minimum for the large black hole.   The van de Waals type transition between the small and large black holes occurs at  $T_c= 0.28475$. 

The effective Gibbs free energy is still given by (\ref{G2})  in the presence of an additive noise.
For multiplicative noise, one finds from (\ref{Ge}) 
\bea
G(r,T) &=& \int^r_c\,\frac{1}{g^2}\left(\frac32 r^2-2\pi T r +\frac12 -\frac{q^2}{2 r^2}\right)dr \n\\
&&+ \frac{k}{L^2}T \left(2-\lambda\right) \ln g(r).
\label{Ge2}
\eea
Phase structures and transitions can then be studied with this effective Gibbs function for different noise profile $g(r)$ as in the previous section. 

As before we  consider only $g(r)=1+a - sign(a) \sqrt{a^2-(r-|a|)^2}$ for $r<|a|$, and $g(r)=1$ for $r>|a|$ ($a>-1$).
As previously noted, black hole solutions of the Einstein equation exist  only when the extrema of $G(r,T)$ occur in the region where $g(r)=1$, i.e., in $r>|a|$.  Otherwise the extrema are transient states. Thus phase structure of the thermodynamic system depends very much on how the locations of the extrema vary as temperature changes.

Fig.\,8 shows $G(r, T_c)$ for $q=1/10$ and different values of $a$ at $T_c=0.28475$ (with $L=1, c=0.005$).  The constant $c$ is chosen such that $G(0.05, T)=0$.  We see that the system is in the small black hole or transient phase when $-1<a<0$,  and has already transited to the large black hole phase when $a>0.2$. This conforms to the previous observation that strong noise level disfavors black hole formation.

To have a better picture of the possible phase structures of the system, we show in Fig.\,9 plots of $G(r,T)$ for different values of $a$ and $T$.  For $a<0$ one generally have a first-order transition between metastable thermal transient phase and stable large black hole.
For $a>0$ we have the following typical situations.  For small enough $a$ such as $a=0.1$, the local minimum  is in the $g=1$ region near transition temperature, so  one sees a first-order transition between stable small black hole and large black hole. For moderate values of $a$ such as $a=0.2$, it would be a first-order transition between transient state to large black hole. Lastly, for larger $a$ such as $a=0.5$, the transition is basically a second-order one between the transient state and the large black hole state.   In Fig.\,10 we show the phase structures for the case $a=0$ and $a=0.2$ in terms of the $G-T$ diagrams.  The phases are similar except that in the case of $a=0.2$, the small black hole phase in the $a=0$ case  is replaced by the thermal transient states.  Also, the first-order transition occurs at a lower critical temperature as noted before.

The discussion in this section concerns mainly how the multiplicative noise affects the first-order  phase transition between small/large RN-AdS black holes.  Such transition is present when the black hole charge is smaller than the critical value, i.e. $q<q_c=1/6$.  In fact the value of $q_c$ can be altered by multiplicative noise.  We check numerically that the change is rather insignificant for small and moderate values of $a>0$, while for large enough $a>0$ there is just second-order transient/large black states for any $q$.  For $-1<a<0$, the critical charge is larger, $q>/1/6$.  This could be understood as follows.  Fig.\,6 indicates that, in the $a=0$ case,  the large black hole states are preferred at larger values of $q$.  As large black hole is disfavored when $a<0$,  a large charge is thus needed to enable formation of large black holes, hence a larger $q_c$.

%%%%%%%%
\section{Summary}
 In this work we have attempted to extend the previous work on Gibbs free energy landscape by considering the effects of multiplicative noise.  An effective free energy function is obtained that can be used to study the modification of the thermodynamic phase  transition of a black hole systems.  
It is realized that it is generally difficult to form black holes under a multiplicative noise, 
 unless the effective and the original free energy become extremal at the same horizon radius.  The  latter situation was modeled by some
theoretical noise profiles which are  monotonically increasing/deceasing functions of the horizon radius.  It is found that stronger noise disfavors the formation of black hole.  Also, with the simple theoretical noise profile discussed here, one already see the emergence of new phase structures.  

%\section*{Acknowledgments}

\leftline{-------------}

The work is supported in part by the Ministry of Science and Technology (MOST)
of the Republic of China under Grants  NSTC 113-2112-M-032-010 and NSTC 114-2112-M-032-005.

%%%%%%%%%%

%%%%%%%%
\newpage
%-----    Fig. 1 ----
\begin{figure}
\includegraphics[width=10cm]{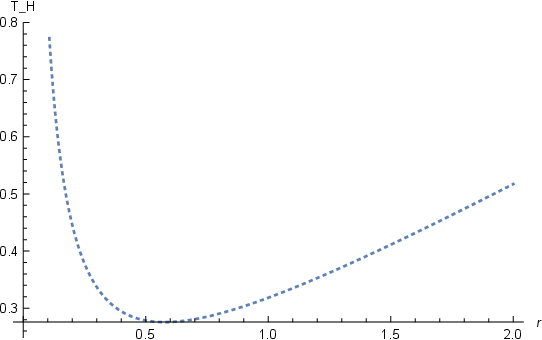}
\caption{Plot of the Hawking temperature $T_H$ in Eq.\,(\ref{data}).}
\label{Fig1}
\end{figure}

%-----    Fig. 2 ----
\begin{figure}
\includegraphics[width=10cm]{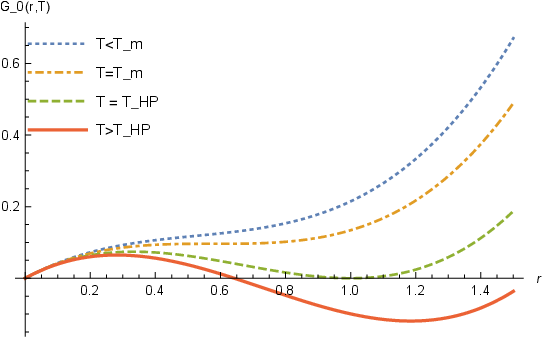}
\caption{Plot of the Gibbs free energy $G_0(r,T)$ for the Schwarzschild-AdS system at different ensemble temperatures. Here $T_m=\sqrt{3}/2\pi=0.27567$ and $T_{HP}=1/\pi=0.31831$.}
\label{Fig2}
\end{figure}

%-----    Fig. 3 ----
\begin{figure}
\includegraphics[width=10cm]{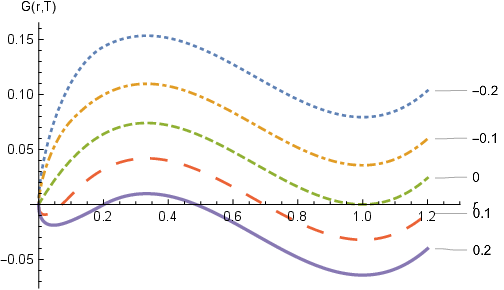}
\caption{Plot of the effective Gibbs free energy $G(r, T_{HP})$ in Eq.\,(\ref{Ge}) for the Schwarzschild-AdS system, labeled by $a$,  at $T_{HP}=1/\pi$ (for $a=0$) with noise profile $g(r)=1+a - sign(a) \sqrt{a^2-(r-|a|)^2}$ for different values of $a$.}
\label{Fig3}
\end{figure}

%------   Fig. 4 ---
\begin{figure}
\includegraphics[width=10cm]{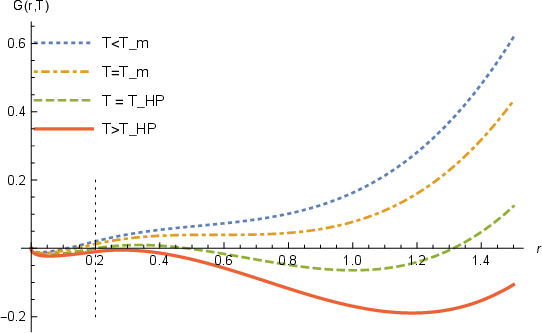}
\caption{Plot of the effective Gibbs free energy $G(r,T)$ for the Schwarzschild-AdS system vs $r$  for different ensemble temperatures with  $g(r)$ defined by $a=0.2$. Here $T_m=\sqrt{3}/2\pi=0.27567$ and $T_{HP}=1/\pi=0.31831$. The noise function $g(r)>1$ differs from the additive noise ($g(r)=1$)  only on the left side of the  vertical dotted line $x=a$. Extremum on the left side of $r=a$ corresponds to thermal transient state,  whilst the extrema on the right side correspond to unstable small/stable large black hole solutions of the Einstein equation.}
\label{Fig4}
\end{figure}

% -----   Fig. 5 ---
\begin{figure}
\includegraphics[width=7cm,height=7cm]{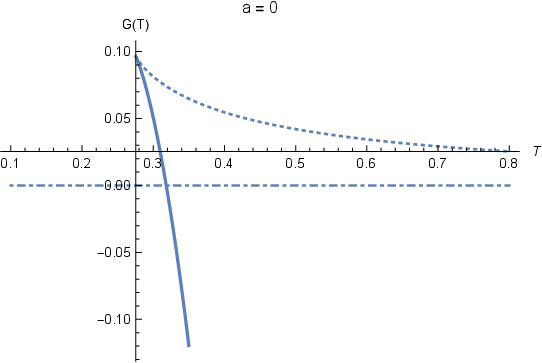}
\hspace{1cm}
\includegraphics*[width=7cm,height=7cm]{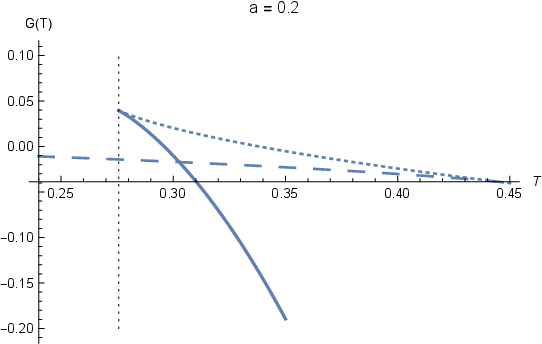}
\caption{Plots of the phase structures of the Schwarzschild-AdS black spacetime in terms of $G-T$  diagrams for the case with the additive noise ($a=0$) and the multiplicative case (for $a=0.2$).  The dotted/solid curves represent unstable small black holes/stable large black hole phases, respectively.  
These two curves bifurcate at $T_m=\sqrt{3}/2\pi=0.27567$ in both plots (shown by the $y$-axis for $a=0$ plot and the dotted vertical line for $a=0.2$ plot). The dashdotted horizontal line shows the thermal AdS radiation phase for $a=0$ case, and the lower long-dashed curve corresponds to the metastable thermal transient phase for $a=0.2$ case.}
\label{Fig5}
\end{figure} 

%-----    Fig. 6 ----
\begin{figure}
\includegraphics[width=10cm]{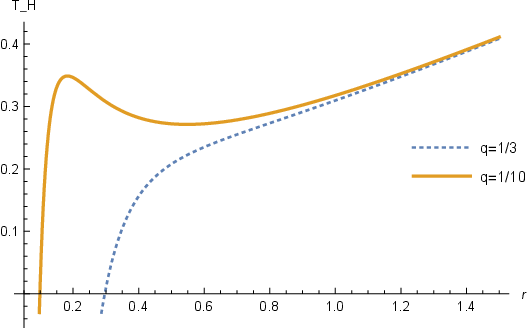}
\caption{Plot of the Hawking temperature $T_H$ in Eq.\,(\ref{Th}) for $q=1/3$ and $1/10$.}
\label{Fig6}
\end{figure}

%-----    Fig. 7 ----
\begin{figure}
\includegraphics[width=10cm]{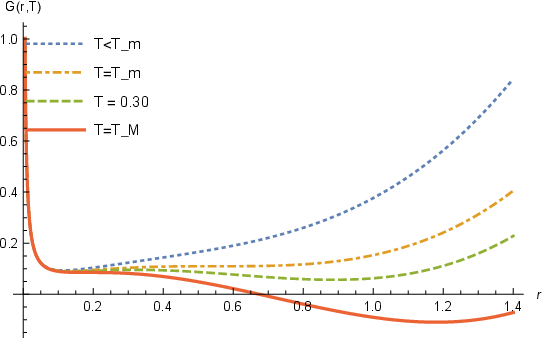}
\caption{Plot of the Gibbs free energy $G_0(r,T)$ for the RN-AdS system vs $r$ with $q=1/10$ for different ensemble temperatures.
Here $T_m=0.27120$ and $T_M=0.34869$.}
\label{Fig7}
\end{figure}

%-----    Fig. 8 ----
\begin{figure}
\includegraphics[width=10cm]{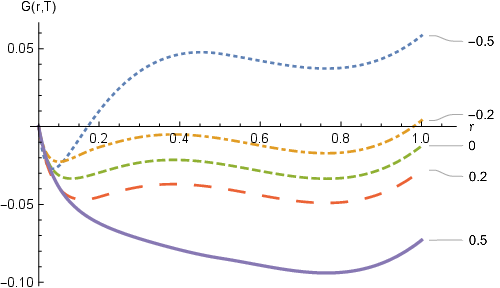}
\caption{Plot of the effective Gibbs free energy $G(r, T_c)$ for the RN-AdS system, labeled by $a$, at $T_c=0.28475$ (for $a=0$) with noise profile $g(r)=1+a - sign(a) \sqrt{a^2-(r-|a|)^2}$ for different values of $a$.}
\label{Fig8}
\end{figure}

% ---  Fig. 9 ----
\begin{figure}
\centering
\includegraphics*[width=7cm,height=7cm]{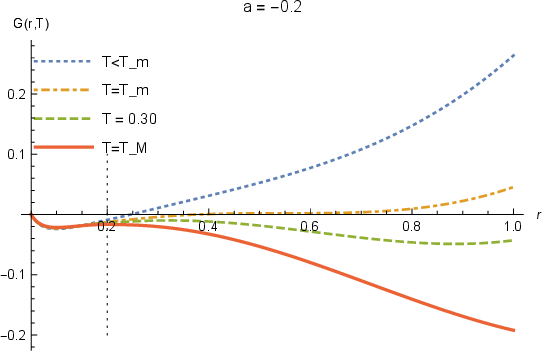}\hspace{1cm}
\includegraphics*[width=7cm,height=7cm]{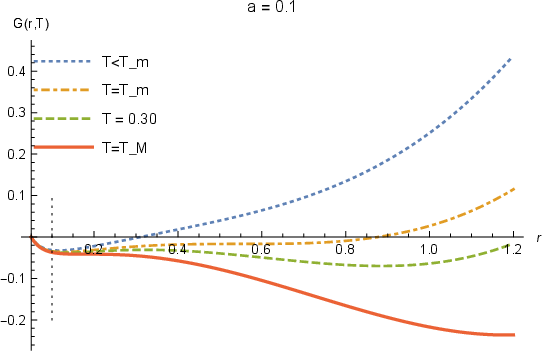}\\
\includegraphics*[width=7cm,height=7cm]{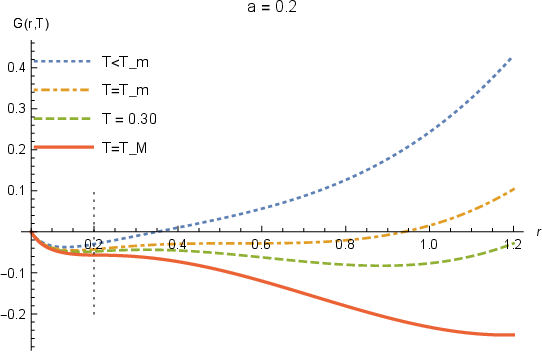}\hspace{1cm}
\includegraphics*[width=7cm,height=7cm]{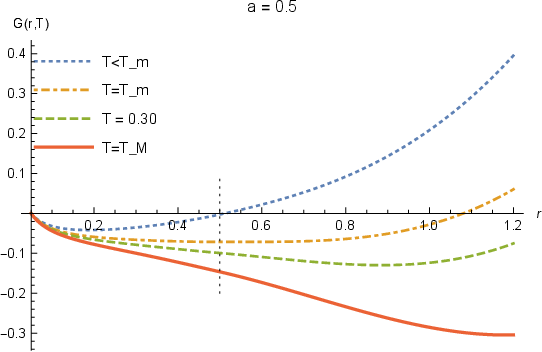}
\caption{Plots of the effective Gibbs free energy $G(r, T)$ vs $r$ for the RN-AdS system for different values of $a$. 
These plots illustrate various types of thermodynamic phase transition between black hole spacetime phases, depending on whether the first local minimum appears on the left or the right side of $r=a$ (indicated by the dotted vertical line). For $a=\pm 0.2$: a first-order transitions between metastable thermal transient phase and stable large black holes.  For $a=0.1$:  a first-order transition between stable small black hole and large black hole.  For $a=0.5$: a second-order transition between transient state and large black hole.  Transition to the large black hole is favored with larger value of $a$.}
\label{Fig10}
\end{figure}

% -----   Fig. 10---
\begin{figure}
\includegraphics*[width=10cm]{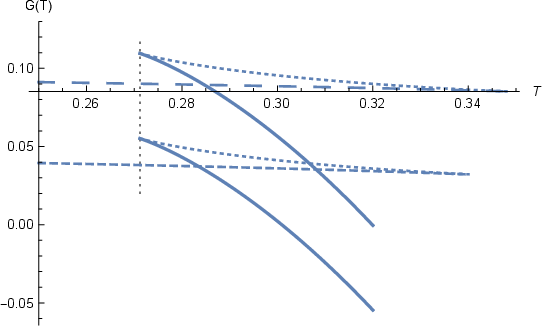}
\caption{Plots of the phase structures of RN-AdS the black holes states in terms of $G-T$  diagrams.  The upper set depicts the case with the additive noise ($a=0$), and the lower set the multiplicative case (for $a=0.2$) (values of $G(r,T)$ shifted for a clearer visual comparison with $G_0(r,T)$).  The dotted/solid curves represent unstable small black holes/stable large black hole phases, respectively.  The long-dashed nearly  horizontal line shows the stable small black hole phase for $a=0$, and the short-dashed line  corresponds to the metastable thermal transient phase for $a=0.2$.  The dotted vertical line indicates the ensemble temperature $T_m=0.27120$, above which  the small and large black hole states bifurcate.}
\label{Fig11}
\end{figure}


\newpage 
\begin{thebibliography}{99}

%%%%%%%%

\bibitem{B}
J.D. Bekenstein, Black holes and entropy, Phys. Rev. D 7 (1973) 2333.

\bibitem{H}
 S. W. Hawking, Particle Creation by Black Holes, Comm. Math. Phys. 43 (1975)199 (1975) [erratum: Comm. Math. Phys. 46 (1976) 206].

\bibitem{BBH}
J.M. Bardeen, B. Carter, and S. Hawking, The four laws of black hole mechanics, Comm. Math. Phys. 31 (1973) 161.

\bibitem{W}
R. M. Wald, The thermodynamics of black holes, Living Rev. Rel. 4 (2001) 6. 

\bibitem{C}
S.Carlip, Black Hole Thermodynamics, Int. J. Mod. Phys. D 23 (2014) 1430023.

\bibitem{HP}
S. Hawking and D.N. Page, Thermodynamics of black holes in anti-de Sitter space, 
Commun. Math. Phys. 87 (1983) 577.

\bibitem{D}
P. C. W. Davies, Thermodynamic Phase Transitions of Kerr-Newman Black Holes in de Sitter Space, 
Class. Quant. Grav. 6 (1989) 1909.

\bibitem{P}
D. Pavon, Phase transition in Reissner-Nordstrom black holes, Phys. Rev. D 43 (1991) 2495.

\bibitem{K1}
O. Kaburaki, Critical behavior of extremal Kerr-Newman black holes, Gen. Rel. Grav. 28 (1996) 843.

\bibitem{CEJM}
A. Chamblin, R. Emparan, C. Johnson, and R. Myers, Charged AdS black holes and catastrophic holography, Phys. Rev. D 60 (1999) 064018.

\bibitem{Wu}
X. N. Wu, Multicritical phenomena of Reissner-Nordstr\"om anti-de Sitter black holes, Phys. Rev. D 62 (2000) 124023.

\bibitem{S}
Y. Sekiwa, Thermodynamics of de Sitter Black Holes: Thermal Cosmological Constant,
Phys.Rev. D73 (2006) 084009.

\bibitem{K2}
D. Kastor, S. Ray, and J. Traschen, Enthalpy and the mechanics of AdS black holes, Class. Quant. Grav. 26 (2009) 195011.

\bibitem{D1}
B. P. Dolan, The cosmological constant and the black hole equation of state, Class.
Quant. Grav. 28 (2011) 125020,

\bibitem{M1}
D. Kubiznak and R.B. Mann, P-V criticality of charged AdS black holes, JHEP 07 (2012) 033.

\bibitem{D2}
B. P. Dolan, Pressure and volume in the  first law of black hole thermodynamics, Class. Quant. Grav. 28 (2011) 235017.


\bibitem{M2}
D. Kubiznak, R. B. Mann, and M. Teo, Black hole chemistry: thermodynamics with Lambda, 
Class. Quant. Grav. 34 (2017) 063001.

\bibitem{LW1}
R. Li and J. Wang, Thermodynamics and kinetics of Hawking-Page phase transition, Phys. Rev. D 102 (2020) 024085.

\bibitem{LW2}
R. Li, K. Zhang, and J. Wang, Thermal dynamic phase transition of Reissner-Nordstr\"om Anti-de Sitter black holes on free energy landscape, JHEP 10 (2020) 090.

\bibitem{LW3}
R. Li and J. Wang, Generalized free energy landscape of a black hole phase transition, Phys. Rev. D 106 (2022) 106015.

\bibitem{X1}
Z.-M. Xu, B. Wu, and W.-L. Yang, van der Waals  fluid and charged AdS black hole in the Landau theory, Class. Quant.
Grav. 38 (2021) 205008.

\bibitem{X2}
Z.-M. Xu, Fokker-Planck equation for black holes in thermal potential, Phys. Rev. D 104 (2021) 104022.

\bibitem{Y}
S.-J. Yang, R. Zhou, S.-W. Wei, and Y.-X. Liu, 
Kinetics of a phase transition for a Kerr-AdS black hole on the free-energy landscape, 
Phys. Rev. D 105 (2022) 084030.

\bibitem{LW4}
R. Li and J. Wang, Kinetics of Hawking-Page phase transition with the non-Markovian effects, JHEP 05 (2022) 128.

\bibitem{LW5}
C. Liu, R. Li, K. Zhang, and J. Wang, Generalized free energy and dynamical state transition of the dyonic AdS black hole in the grand canonical ensemble, JHEP 11 (2023) 068.

\bibitem{WLL}
W.-Y. Wu, Z. Luo, and J. Li, Thermodynamic phase transition of AdS black holes in massive gravity on free energy landscape, Int. J. Theor. Phys. 63 (2024) 177.

\bibitem{X3}
Z.-M. Xu, B. Wu, and W.-L. Yang, Rate of the phase transition for a charged anti-de Sitter black hole, Sci. China-Phys.
Mech. Astron. 66 (2023) 240411.

\bibitem{A1}
Mohammad Ali S. Afshar, Saeed Noori Gashti, Mohammad Reza Alipour, and Jafar Sadeghi,
Kramers escape rate and phase transition dynamics in AdS black holes
arXiv:2404.17849.

\bibitem{X4}
C. Ma, P.-P. Zhang, B. Wu, Z.-M. Xu,
The Kramers escape rate of phase transitions for the 6-dimensional Gauss-Bonnet AdS black hole with triple phases,
Phys. Lett. B 861 (2025) 139282.
%arXiv:2407.20512.

\bibitem{A2}
Mohammad Ali S. Afshar, Saeed Noori Gashti, Mohammad Reza Alipour, and Jafar Sadeghi,
Overcoming Barriers: Kramers escape rate analysis of metastable dynamics in first-order multi-phase transitions,
J. High Energ. Phys. (2025) 163.
%doi.org/10.1007/JHEP11(2025)163
%arXiv:2506.07074.

\bibitem{R}
H. Risken, The Fokker-Planck Equation, 2nd. ed., Springer-Verlag, Berlin, 1996.

\end{thebibliography}
\end{document}